\documentclass{ws-procs9x6}
\usepackage{graphics,latexsym}
\newcommand{\bea}{\begin{eqnarray}}
\newcommand{\eea}{\end{eqnarray}}
\newcommand{\calm}{{\mathcal{M}}}

\newcommand{\del}{\partial}
\renewcommand{\vec}[1]{\ensuremath{\mathbf{#1}}}
\begin{document}
\title{Nonequilibrium dynamics in scalar hybrid models%
}
\author{J\"urgen Baacke \and Andreas Heinen}
\address{Institut f\"ur Physik, Universit\"at Dortmund,
D-44221 Dortmund , Germany}
\maketitle
\abstracts
{
We study by numerical simulations the transition from the metastable
``false vacuum'' to the broken symmetry phase in the preheating stage 
after cosmic inflation in a scalar hybrid model.
We take quantum fluctuations and their back reaction into account 
by applying a one-loop bubble-resummation.\cite{Baacke:2003bt}
}

\section{Introduction}
At present there is a wide class of inflationary 
models, having in common the aim of explaining 
various kinds of cosmological puzzles\cite{Lyth:1998xn}. 
One of the consecutive questions raised is how inflation has ended. 
In a scenario called ``Hybrid inflation''\cite{Linde:1990gz} 
the inflationary expansion ends by a phase transition
from a metastable ``false vacuum'' to the true, i.e. broken 
symmetry vacuum. The subsequent phase of (p)reheating\cite{preheating} 
after inflation is characterized by a phase of abundant 
particle production due to spinodal amplification of quantum modes 
and due to parametric resonance.
We will study some aspects of this phase in the following.


\section{Effective action and renormalized equations of motion}
The Lagrangian for the Hybrid model is given by
\bea
\mathcal{L}&=&\frac{1}{2}\del_\mu\Phi\del^\mu \Phi+\frac{1}{2}\del_\mu
X\del^\mu X-\frac{1}{2}m^2\Phi^2-\frac{1}{2}g^2\Phi^2 X^2
-\frac{\lambda}{4}(X^2-v^2)^2 \ .
\eea
The field $X$ has an effective mass square 
$m^2_X(\Phi)=-\lambda v^2 + g^2 \Phi^2$ and becomes unstable 
if the absolute value of $\Phi$ 
is lower than $\Phi_\mathrm{c}=\frac{v\sqrt{\lambda}}{g}\label{eq:phi_c}$ 
(spinodal region).
Both quantum fields $\Phi$ and $X$ are assumed to have classical 
expectation values, i.e.
$\left\langle \Phi\right\rangle=\phi$ (inflaton field) and 
$\left\langle X\right\rangle=\chi$ (symmetry breaking 
field).

We will use here the so-called \emph{two-particle point-irreducible} 
(2PPI) effective action (EA) formalism.
\cite{Verschelde:1992bs,Baacke:two-loop-2PPI}
The renormalized equations of motion (EOM) that describe 
the nonequilibrium dynamics of classical mean fields and their
quantum fluctuations, all coupled to each other, 
are derived from the EA functional
\bea
\Gamma[\phi,\chi,\Delta_{\{\phi\phi,\phi\chi,\chi\chi\}}]
&=&S[\phi,\chi]
+
\frac{g^2}{2}\int d^Dx\left(\Delta_{\phi\phi}\Delta_{\chi\chi}
+2\Delta_{\phi\chi}^2\right)\\[-0.05in]
&&+\frac{3 \lambda}{4} \int d^Dx \Delta_{\chi\chi}^2
+\Gamma^\mathrm{2PPI}[\phi,\chi,\calm^2_{\{\phi\phi,\phi\chi,\chi\chi\}}]
\ ,\nonumber \\[-0.05in] 
\Gamma^\mathrm{2PPI}[\phi,\chi,\calm^2_{\{\phi\phi,\phi\chi,\chi\chi\}}] 
&=& \raisebox{-0.15cm}{\includegraphics[width=0.5cm]{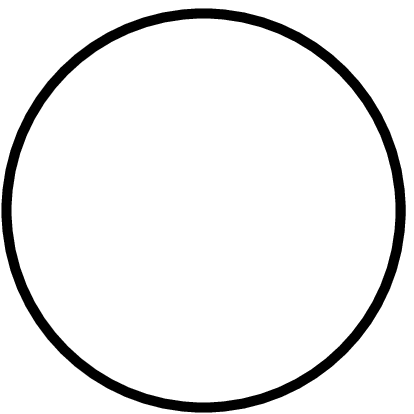}}
+\raisebox{-0.15cm}{\includegraphics[width=1cm]{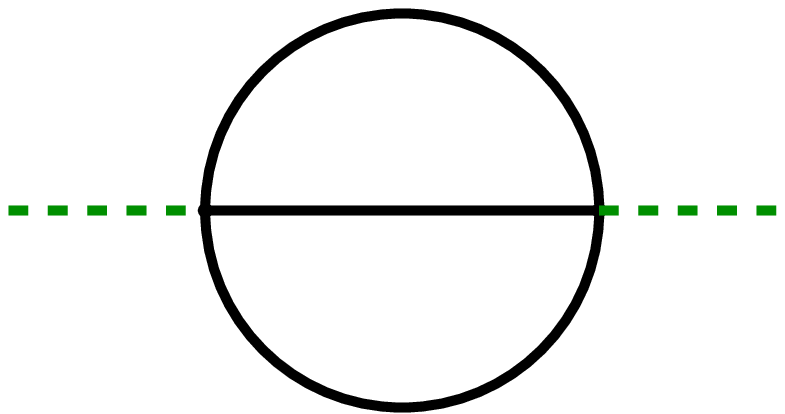}}
+\raisebox{-0.15cm}{\includegraphics[width=0.5cm]{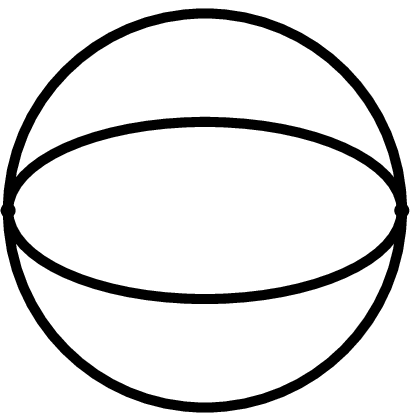}}+
\raisebox{-0.15cm}{\includegraphics[width=0.6cm, angle=90]{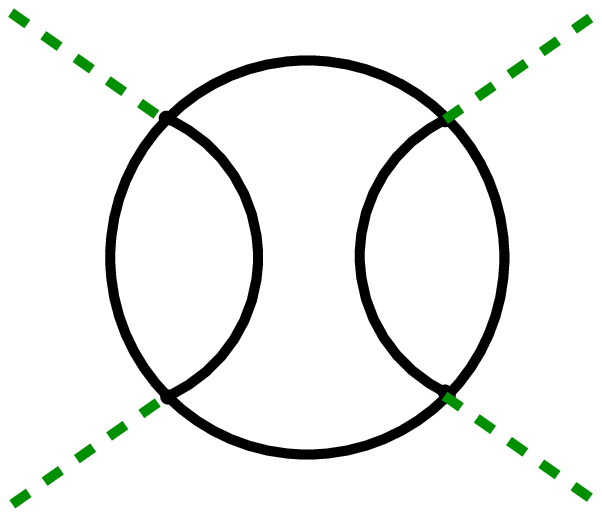}}
+\ldots \label{eq:Gamma2PPI}\ .
\eea
The classical EOM follow from 
$\frac{\delta \Gamma}{\delta \phi}=0$ and 
$\frac{\delta \Gamma}{\delta \chi}=0$, 
while the gap equations are derived from 
$\frac{\delta \Gamma}{\calm^2_{ij}}=0$ with $ij=\phi\phi,\chi\chi$ 
or $\phi\chi$.
The propagator $G$ fulfills a constraint equation
\bea
(G^{-1})_{ij}(x,x')
=i\left(\square \delta_{ij}+\calm^2_{ij}(x)\right)\delta^{(D)}(x-x')
\label{eq:EOM-G} \ .
\eea
It is dressed by a resummation of local self-energy insertions.
Assuming spatial homogeneous fields, $G_{ij}(t,t';\vec{p})$ 
can be decomposed in momentum space into mode functions via 
$G_{ij}(t,t';\vec{p})=\sum_{\alpha=1}^2\frac{1}{2\omega_\alpha}
f_{i}^\alpha(t;p)f_{j}^{*\,\alpha}(t';p)$. 
The quantities $\omega_\alpha$ ($\alpha=1,2$) have to be determined self 
consistently at the initial time ($t=0$). 
For practical calculations one has to truncate the infinite series in 
Eq.(\ref{eq:Gamma2PPI}). Here we will only keep the one-loop term
which leads to a summation of
bubble diagrams.\footnote{Equivalently one can take the 
\emph{local part} of a two-loop-2PI approximation.} It is given by
$\Gamma^{(1)}[\calm^2_{\phi\phi},\calm^2_{\chi\chi},\calm^2_{\phi\chi}]
=\frac{i}{2}\mathrm{Tr}\ln[G^{-1}]$.
The tadpole insertions derived from $\Gamma^{(1)}$ denote
\bea
\Delta_{ij}^{(1)}(t)&=&
-2\frac{\delta\Gamma^{(1)}}{\delta \calm^2_{ij}(t)}
=\sum_{\alpha=1}^2\int\frac{d^{D-1}p}{(2\pi)^{D-1}}
\frac{1}{2\omega_\alpha}\mathrm{Re}
\left[f_i^\alpha(t;p)f_j^{\alpha\,*}(t;p)\right] \ .
\eea
These quantities have quadratic and logarithmic divergencies and have to
be renormalized. It turns out\cite{Baacke:2003bt} that 
the EA is renormalized by a simple vacuum counter 
term $\delta E_\mathrm{vac}=
-\delta\xi[(\calm^2_{\phi\phi})^2+2(\calm^2_{\phi\chi})^2
+(\calm^2_{\chi\chi})^2]$, where in dimensional regularization  
$\delta\xi=-\frac{1}{64\pi^2}(\frac{2}{\epsilon}-\gamma+1+\ln
4\pi)$. The vacuum counter term $\delta E_\mathrm{vac}$ indeed reflects a 
resummation of the standard counterterms derived 
from a counterterm Lagrangian.\cite{Verschelde:1992bs}
The renormalized gap equations for the variational parameters 
$\calm^2_{ij}$ take the form
\bea
\calm^2_{\mathrm{R},\,\phi\phi}(t)&=&
m^2+g^2\left[\chi^2(t)+\Delta_{\chi\chi}(t)\right]
-4g^2\, \delta\xi\calm^2_{\mathrm{R},\chi\chi}(t)\label{eq:gap1}\ ,\\
\calm^2_{\mathrm{R},\,\chi\chi}(t)&=&
-\lambda v^2+g^2\left[\phi^2(t)+\Delta_{\phi\phi}(t)\right]
+3\lambda\left[\chi^2(t)+\Delta_{\chi\chi}(t)\right]\nonumber\\
&&-4g^2 \,
\delta\xi\calm^2_{\mathrm{R},\,\phi\phi}(t)-12\lambda\,
\delta\xi\calm^2_{\mathrm{R},\,\chi\chi}(t)\label{eq:gap2} \ ,\\
\calm^2_{\mathrm{R},\,\phi\chi}(t)
&=&2g^2\left[\phi(t)\chi(t)+\Delta_{\phi\chi}(t)\right]
-8g^2\,\delta\xi\calm^2_{\mathrm{R},\,\phi\chi}(t) \label{eq:gap3}\ .
\eea
The final set of gap equations, with all finite parts from dimensional 
regularization, is rather lengthy\cite{Baacke:2003bt} and
we don't write it down here. However, one can see from 
Eqs.~(\ref{eq:gap1})-(\ref{eq:gap3}) that the gap equations form a 
($3\times 3$) system of linear equations, whose coefficient matrix  
has to be diagonalized by a time-independent rotation matrix. This 
fixes the various mass and coupling constant counterterms, obtained 
from a standard counterterm Lagrangian, in a non-perturbative way.

After all the classical EOM are given by
\bea 
\ddot{\phi}(t)+\calm^2_{\mathrm{R},\,\phi\phi}(t)\phi(t)
+\calm^2_{\mathrm{R},\,\phi\chi}(t)\chi(t)-2g^2\chi^2(t)\phi(t)=0
&&\label{eq:EOM-phi}\!\! , \\
\ddot{\chi}(t)+\calm^2_{\mathrm{R},\,\chi\chi}(t)\chi(t)
+\calm^2_{\mathrm{R},\,\phi\chi}(t)\phi(t)-2\lambda \chi^3(t)
-2g^2\phi^2(t)\chi(t)=0&&\! 
\label{eq:EOM-chi} .
\eea
Inserting the decomposition of $G_{ij}(t,t;\vec{p})$ into 
mode functions $f_i^\alpha(t,p)$ in Eq.~(\ref{eq:EOM-G})  
the system of EOM for the $f_i^\alpha(t,p)$ denotes explicitly as
\bea
\ddot{f}_\phi^\alpha (t;p)+\vec{p}^2
f_\phi^\alpha(t;p)+\,\calm^2_{\mathrm{R},\,\phi\phi}(t) f_\phi^\alpha(t;p)
+\calm^2_{\mathrm{R},\,\phi\chi}(t) f_\chi^\alpha(t;p)
=0\label{eq:EOM-f_phi}&&\!\! ,\\
\ddot{f}_\chi^\alpha (t;p)+\vec{p}^2
f_\chi^\alpha(t;p)
+\,\calm^2_{\mathrm{R},\,\chi\chi}(t) f_\chi^\alpha(t;p)
+\calm^2_{\mathrm{R},\,\phi\chi}(t) f_\phi^\alpha(t;p)=0&& 
\label{eq:EOM-f_chi}\!\! .
\eea

\section{Results}
The EOM (\ref{eq:EOM-phi})-(\ref{eq:EOM-f_chi}) are solved numerically. 
After renormalization the momentum integrals are finite and 
are carried out using a non-equidistant momentum discretization.
In order to simulate the phase transition at the end of inflation 
we choose the initial amplitudes $\phi(0)> \phi_\mathrm{c}$ and 
$\chi(0)\approx 0$. In particular we take $\phi(0)=1.2$ 
and $\chi(0)=10^{-7}$. The other parameters are here
$m^2=0$, $v=1$, $g=2\lambda$, $\lambda=1$. 

\begin{figure*}[htbp]
\begin{center}
\includegraphics[width=0.91\linewidth]{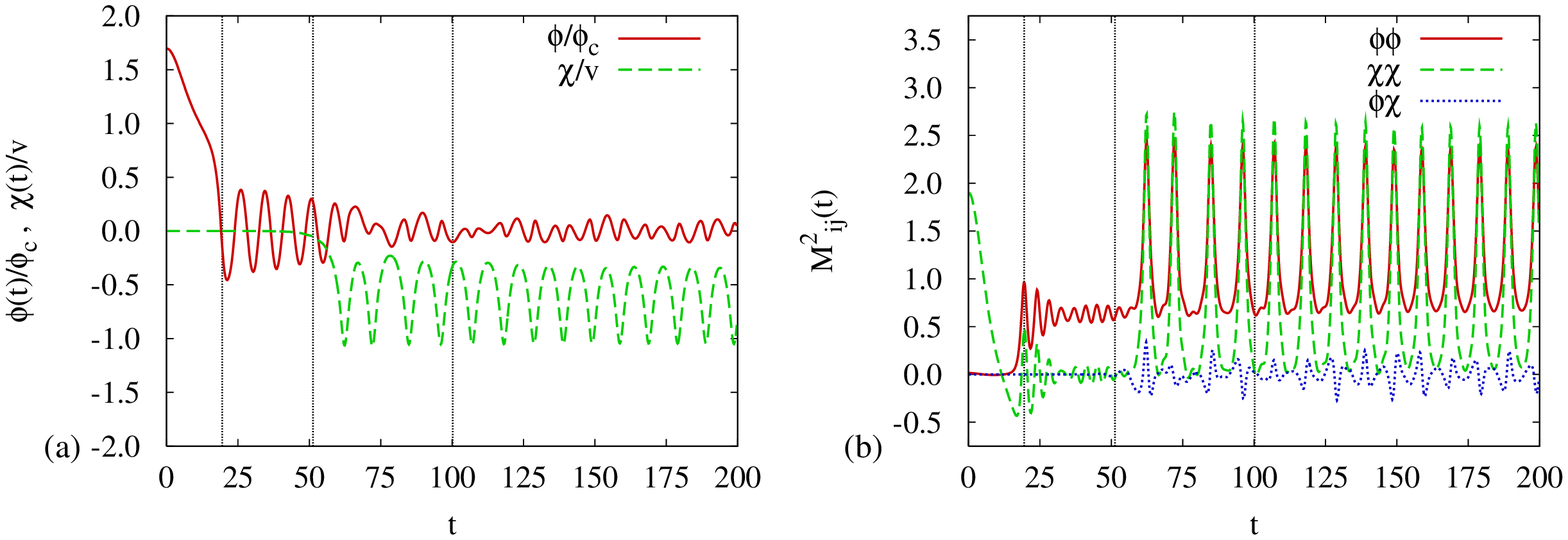}
\end{center}
\vspace{-0.15in}
\caption{Time evolution of 
(a) the classical fields 
$\phi(t)/\phi_\mathrm{c}$ (solid line) 
and $\chi(t)/v$ (dashed line), 
(b) the effective masses $\calm^2_{ij}(t)$ with $ij={\phi\phi}$ 
(solid line), $ij={\chi\chi}$ (dashed line) and $ij={\phi\chi}$ 
(dotted line); 
the vertical dotted lines indicate 
the times $19.5$, $51.3$ and $100.2$; parameters: $g^2=2\lambda$, 
$\phi(0)=1.2$, $\chi(0)=1.0\times 10^{-7}$, $m^2=0$, 
$\lambda=1$ and $v^2=1$.}
\label{fig:phi-M2}
\end{figure*}

\begin{figure*}[htbp]
\begin{center}
\vspace{-0.15in}
\includegraphics[width=0.90\linewidth]{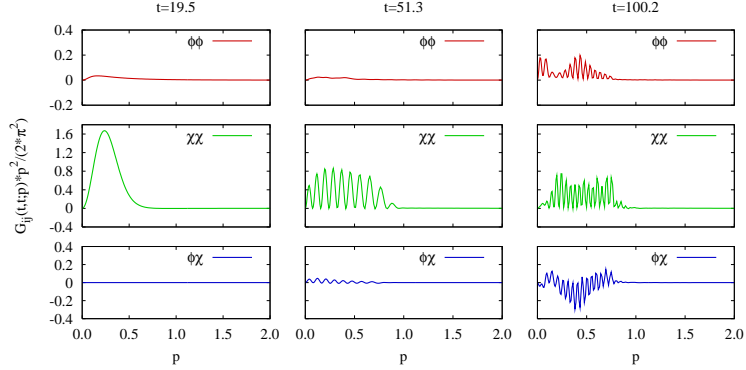}
\end{center}
\vspace{-0.15in}
\caption{Momentum spectra $G_{ij}(t,t;\vec{p})p^2/(2\pi^2)$ 
for the simulation in Fig.~\ref{fig:phi-M2} at the times $t=19.5$ 
(left column), $t=51.3$ (middle column) and $t=100.2$ (right column).}
\label{fig:spectra}
\end{figure*}

The time evolution of 
the classical fields $\phi$ and $\chi$ and of the variational masses is 
displayed in Fig.~\ref{fig:phi-M2}. One observes three 
time regimes: 
(i) the first regime is characterized 
by a slow-roll of the field $\phi$, 
where quantum fluctuations are almost negligible;
(ii) in the spinodal region the mass $\calm^2_{\chi\chi}$ becomes negative 
(Fig.~\ref{fig:phi-M2}b). Gaussian fluctuations build up
(left column in Fig.~\ref{fig:spectra}) and drive $\calm^2_{\chi\chi}$ 
back to positive values. Along the way the amplitude of $\chi$ 
growths exponentially; 
(iii) at later times the momentum spectra display several spikes 
(see e.g. the right column in Fig.~\ref{fig:spectra}) 
and the symmetry breaking field oscillates around a nonzero value 
with a relatively constant amplitude.
The details in the dynamics 
depend on the parameters in the model while the characteristic 
features are very similar.\cite{Baacke:2003bt}

\begin{figure}[htbp]
\begin{minipage}{0.52\linewidth}
  \centering
  \includegraphics[width=0.94\linewidth]{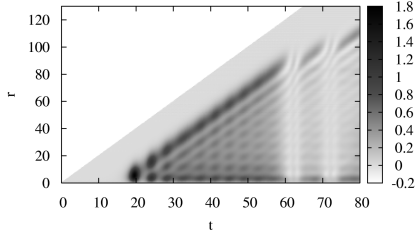}
\vspace{-0.15in}
\end{minipage}
\hfill
\begin{minipage}{0.47\linewidth}
  \centering
  \caption{\newline Correlation function $rC_{\chi\chi}(r,t)=
r\int\frac{d^3p}{(2\pi)^3}e^{i\vec{p}\cdot\vec{r}}
 \sum_\alpha \frac{1}{2\omega_\alpha} 
\mathrm{Re}[f_\chi^\alpha f_\chi^{*\,\alpha}]$ 
for the simulation in Fig.~\ref{fig:phi-M2}. 
The propagation velocity is $r/t=2$ (twice the speed of light).}
  \label{fig:C_chichi}
\end{minipage}
\end{figure}

In Fig.~\ref{fig:C_chichi} we display the time and space evolution 
of correlations between the fluctuations. One can see that spatial 
correlations build up in the spinodal region and propagate by twice the 
speed of light. The correlations decrease once the field $\chi$ starts 
to oscillate around a non-zero expectation value.

\begin{figure}[htbp]
\begin{minipage}{0.51\linewidth}
  \centering
\vspace{-0.1in}
  \includegraphics[width=0.88\linewidth]{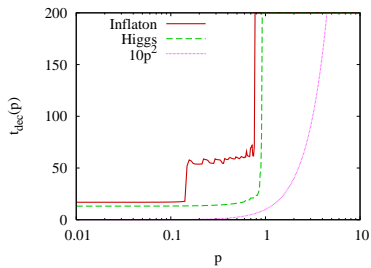}
\vspace{-0.15in}
\end{minipage}
\hfill
\begin{minipage}{0.46\linewidth}
  \caption{De\-coherence time $t_\mathrm{dec}(p)$ for which a given mode 
$p$ becomes ``classical''  
for the simulation in Fig.~\ref{fig:phi-M2}; 
the solid line represents the inflaton  and the 
dashed line the Higgs modes; the dotted line corresponds to 
$t_\mathrm{dec}\propto p^2$.}
  \label{fig:tdec-p}
\end{minipage}
\end{figure}

The decoherence time $t_\mathrm{dec}(p)$  
for which a given mode becomes 
classical is defined by $|F_{ij}(t_\mathrm{dec},\vec{p})|=1$
with $F_{ij}(t,\vec{p})=\mathrm{Im}[\sum_\alpha \frac{f_i^{\alpha *}
\dot f_j^\alpha}{2\omega_\alpha}]$. The modes right to the 
curves in Fig.~\ref{fig:tdec-p} never become classical. 
The functional dependence is not proportional to $p^2$, as found 
in models with a slow quench where 
$\calm^2_{\chi\chi}(t)\propto (t_0-t)$.

\vspace{-0.2in}
\section{Conclusions}
We have analyzed by numerical simulations the transition from the 
metastable phase to the broken symmetry phase in the Hybrid model. 
We observed that 
the back reaction of quantum fluctuations plays an important role and 
leads, e.g., to a dynamical stabilization in the spinodal region. 
The approximation used here provides reliable information on the 
nonequilibrium evolution at earlier times. However, 
one has to improve further in order to cure, e.g. 
the lack of dissipation at later times. 
While there are various improvements, 
once scattering is taken into account via non-local two-loop 
diagrams, even in the 2PPI resummation scheme 
\cite{Baacke:two-loop-2PPI}, one 
ultimately wants to achieve a universal late time behavior, 
i.e. reaching quantum thermalization. The full treatment of the 
reheating phase remains a desirable goal for future investigations.

\vspace{0.05in} 


AH thanks the GK 841 for partial financial support.

\end{document}